\documentclass[pra,aps,twocolumn,showpacs,amsmath,amssymb,floatfix,10pt]{revtex4-1}
\input{epsf}

\usepackage{graphicx}
\usepackage{dcolumn}
\usepackage{bm}

\begin{document}
\title{Particle-hole entanglement of ultracold atoms in an optical lattice}
\author{H. T. Ng}
\affiliation{Center for Quantum Information, Institute for Interdisciplinary Information Sciences, Tsinghua University, Beijing 100084, P. R. China}

\date{\today}

\begin{abstract}
We study the ground state of two-component bosonic atoms in a one-dimensional
optical lattice. By applying an external field to the atoms 
at one end of lattice, the atoms are transported and becomes localized at that site.
The holes are then created in the remaining sites. The particle-hole superpositions 
are produced in this process.  
We investigate the entanglement entropy between the atoms in the two different 
parts of a lattice. A large degree of particle-hole entanglement is generated in the ground state. 
The particle-hole quantum correlations can be probed by the 
two-site parity correlation functions. The transport properties
of the low-lying excited states are also discussed.
\end{abstract}


\maketitle

\section{Introduction}
Ultracold atoms in optical lattices
provide a ground for studying 
quantum many-body systems  \cite{Weitenberg}.
A wide range of interaction parameters
of ultracold atomic systems can be tuned \cite{Greiner}
and remarkable detection techniques
have been demonstrated \cite{Bakr}. For example,
a quantum phase transition \cite{Greiner} from a superfluid
to a Mott insulator has been shown in 
ultracold bosons in an optical lattice.
Also, high-resolution single-site imaging 
has been used for probing the Bose-Hubbard
model \cite{Bakr2}.

Recently, particle-hole correlations
have been directly observed \cite{Endres} in a bosonic 
Mott insulator. Indeed, correlated particle-hole pairs 
are virtual excitations of a Mott
insulator with weak tunnel couplings. 
Apart from this, the propagation speed of 
particle-hole correlations of ultracold atoms \cite{Cheneau} 
has been experimentally studied in a one-dimensional (1D) lattice.
Such speed of correlations is a fundamental
property of the dynamics of a many-body system \cite{Lieb}.

\begin{figure}[ht]
\centering
\includegraphics[height=3.8cm]{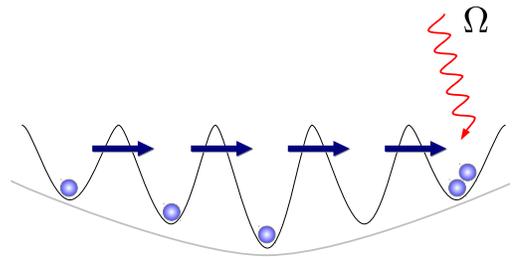}
\caption{ \label{fig1} (Color online) Atoms in a 1D optical lattice 
with a harmonic confinement. 
The local field is individually applied to one end of the lattice
with the coupling strength $\Omega$.}
\end{figure}

In this paper, we consider the ultracold 
bosonic atoms to be trapped in a 1D lattice 
as shown in Fig.~\ref{fig1}, 
where a laser field is individually applied to 
atoms at one end of lattice. Here we consider
on-site interactions to be
much stronger than the tunnel couplings.
The atoms are transported to the site
where the field is applied. In fact, the quantum transport of 
a double-well Bose-Einstein condensates by using an external field
has been recently discussed \cite{Ng}.

The transport and localization 
of atoms will create holes in the remaining sites
if the system is at unity filling. 
This situation is different to particle-hole states 
of the Mott insulator being observed 
in experiments by Endres {\it et. al} \cite{Endres}.
In their experiments, particle-hole
excitations can be understood from
the first-order perturbation theory \cite{Toth}.
In our case, ``real'' holes can be created
in the lattice by using a local external field.
The holes are created in the exact ground
state. When a single atom is transported, a superposition 
of a number of particle-hole states 
are generated. Therefore, strong particle-hole 
correlations can build up. 
These particle-hole correlations can be detected by 
measuring the two-site parity correlation functions 
\cite{Endres,Cheneau}.

In addition, we study the entanglement \cite{Horodecki}
between atoms in the two different parts of a lattice.
The entanglement entropy is used to 
quantify the degree of entanglement. 
The method of detecting
the entanglement entropy in an optical lattice 
has recently proposed \cite{Abanin,Daley}. 
This quantity can provide
useful information of the non-trivial properties
of the ground state of a many-body system \cite{Vidal}. 
We find that a large degree of particle-hole 
entanglement can be generated when an atom
is transported. The ground state becomes 
highly entangled due to creation of holes. 
This means that \textit{local interactions
between the atoms and an external field can 
give rise to novel behaviours of the ground
state}. 

This paper is organized as follows:
In Sec.~II, we introduce the model of
two-component bosons in
an optical lattice and the interactions 
between atoms and a laser field. 
In Sec.~III, we study the transport
of atoms by applying an external field
to the atoms. We also investigate the two-site
parity correlation functions for detecting the 
particle-hole correlations. In Sec.~IV, we study the
entanglement entropy between two parts of
the lattice. We discuss the adiabatic transition
and the transport properties of 
the low-lying excited states in Sec.~V. 
Finally, we provide a conclusion.

\section{System}
We consider two-component bosonic atoms to be trapped
in a 1D lattice in the presence of a harmonic 
trap as shown in Fig.~\ref{fig1}.  Each atom has the two internal
states, i.e., the upper state $|e\rangle$ and lower state $|g\rangle$, 
respectively. 
This system can be described by the Bose-Hubbard model \cite{Fisher}. 
The Hamiltonian of a system of two-component bosons is written as ($\hbar=1$)
\begin{eqnarray}
H&=&-\sum^{M}_{i=1}(J_ee^\dag_i{e}_{i+1}+J_gg^\dag_ig_{i+1}+{\rm H.c.})+U_{eg}\sum^{M}_{i=1}n^e_in^g_i\nonumber\\
&&+\frac{U_e}{2}\sum^{M}_{i=1}n^e_{i}(n^e_i-1)
+\frac{U_g}{2}\sum^{M}_{i=1}n^g_i(n^g_i-1)\nonumber\\
&&+\sum^{M}_{i=1}(\epsilon^e_in^e_i+\epsilon^g_in^g_i)
\end{eqnarray}
where $e^\dag_i(g^\dag_i)$ and $e_i(g_i)$ are the creation and
annihilation operators of an atom in the state $|e\rangle$($|g\rangle$)  at site $i$,
and $M$ is the total number of sites.
The parameters $J_e(J_g)$ and $U_e(U_g)$ are the strengths of tunnel
couplings and the atom-atom interactions of atoms in the states $|e\rangle(|g\rangle)$,
respectively, and $U_{eg}$ is the strength of inter-component interactions
between the atoms. $\epsilon^e_i$ and $\epsilon^g_i$
are the strengths of harmonic confinement of the excited 
and ground states of the atoms, respectively.  They are 
proportional to the square of the distance of the trap centre from
their positions \cite{Greiner2}.  

We have assumed that the atoms are trapped in the lowest band of the lattice.
In fact, the single-band model
is valid \cite{Pichler} provided that the on-site interaction strengths and
the temperature are much lower than the energy gap
between the first excited band and the lowest band. Apart from
this, we consider the interaction strengths between the two
component bosons to be approximately equal, i.e., 
$U_{ee}{\approx}U_{gg}{\approx}U_{eg}{\approx}U$
and $J_e{\approx}J_g{\approx}J$.
In fact, the scattering
lengths of the two hyperfine spin states of ${}^{87}$Rb
are very similar \cite{Matthews}. 

We consider the atoms to be individually coupled to an external 
laser field of the frequency $\omega$. In the interaction picture, 
the Hamiltonian, describes the interaction between the atoms and the laser 
field, is written as \cite{Ng}
\begin{eqnarray}
H_{\rm I}&=&\Delta\sum^M_{i=1}n^e_{i}+\sum^{M}_{i=1}\Omega_i(e^\dag_ig_i+{\rm H.c.}),
\end{eqnarray}
where $\Delta=\omega-\omega_g$ and $\Omega_i$ are the detuning and 
coupling strength between the laser and atom at site $i$, respectively. $\omega_g$
is the frequency of the atom in the state $|e\rangle$ and the energy
of the ground state $|g\rangle$ is set to be zero. Without loss of generality, 
we consider the external field to be individually applied to the atoms at the 
one end of lattice, namely, site $M$.

\section{Transport and localization}
\begin{figure}[ht]
\centering
\includegraphics[height=5.0cm]{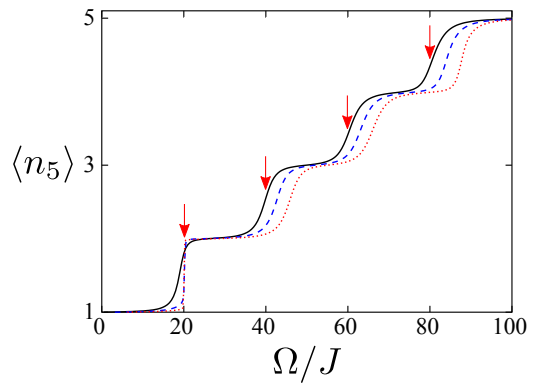}
\caption{ \label{qtfig1} (Color online) Number of atoms
at site 5, $\langle{n_5}\rangle$ versus
coupling strength $\Omega$, for $N=M=5$, $U=20J$ and $\Delta=0$.  
The field is applied to the atoms at site 5. 
The harmonic confinement strengths $\epsilon^e_i=\epsilon^g_i$ are
both equal to $(i-3)^2\epsilon$. The different parameters $\epsilon$ 
are denoted by the different lines: $\epsilon=0$ (black-solid), 
$\epsilon=J$ (blue-dashed) and $\epsilon=2J$ (red-dotted), respectively.
The coupling strengths $\Omega^*_n$ are marked with red arrows.}
\end{figure}

We study the ground state of the coupled system
of the atoms and external field in the strongly
interaction regime, i.e., $U\gg{J}$. We consider
that the average number of atoms in each site is equal to one.
We employ the exact diagonalization method \cite{Zhang}
for numerical simulation.
In Fig.~\ref{qtfig1}, we plot the number of atoms
 at site $M$ versus the coupling
strength $\Omega$ without the harmonic
confinement. The atoms 
can be transferred to that site $M$ by increasing
the coupling strength of the field and atoms.
The atoms are transported stepwise \cite{Ng}
around the specific coupling strengths $\Omega^*_n$ which are given by
\begin{eqnarray}
\Omega^*_n&=&\frac{1}{2}[(2Un+\Delta)^2-\Delta^2]^{1/2}
\end{eqnarray}
The derivation of $\Omega^*_n$ is given in
Appendix A. In the strongly interaction regime, 
a single atom is only allowed to transport \cite{Ng}.
All atoms can be transferred to site $M$ and
becomes localized if $\Omega$ is sufficiently
large.

We also study the transport of atoms in the 
presence of weakly harmonic confinement.
In the same figure, we plot the number of 
atoms at site $M$ as a function of coupling 
strength $\Omega$ for the different strengths
of harmonic confinement.  Similarly, the atoms can be transported
to site $M$. This shows that
the atoms can be transported by
applying an external field even if
the weakly harmonic confinement
is present.

\subsection{Particle-hole correlation}

\begin{figure}[ht]
\centering
\includegraphics[height=5.0cm]{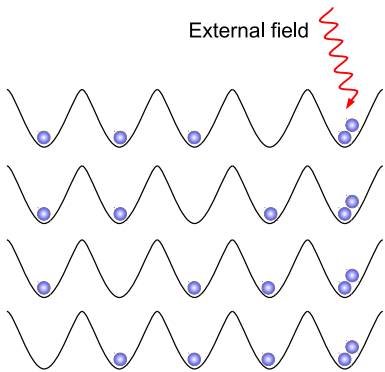}
\caption{ \label{particle_hole1} (Color online) Four different outcomes of particle-hole
states, where the number of sites $M$ and atoms $N$ are 
both equal to 5. The two atoms are localized at the end of the lattice by applying an external
field to the atoms at one end of lattice. This causes a hole in the lattice. There are
four possible outcomes when a hole is created.}
\end{figure}

Since the system is prepared
at unity filling, the localization of
atoms leads to creation of holes in the 
remaining lattice sites. A superposition
of particle-hole states is produced and
therefore particle-hole correlations can
be generated. The possible outcomes of the
system are schematically depicted in Fig.~\ref{particle_hole1}.
We study the particle-hole states between site $M$ 
and site $M-d$ by calculating the
two-site parity correlation function \cite {Weitenberg,Cheneau} as:
\begin{eqnarray}
C(d)&=&|\langle{s_Ms_{M-d}}\rangle-\langle{s_M}\rangle\langle{s_{M-d}}\rangle|,
\end{eqnarray}
where $s_M=e^{i\pi{n_M}}$ is the parity operator 
at site $M$ and $d$ is the index of the number of sites
between two sites.  The quantity $\langle{s_i}\rangle$  depends on
the number of atoms at site $i$. Creation of holes leads to the changes
of the number of atoms in each site.
Therefore, $C(d)$ can indicate 
the particle-hole correlations between the 
two sites.

\begin{figure}[ht]
\centering
\includegraphics[height=5.0cm]{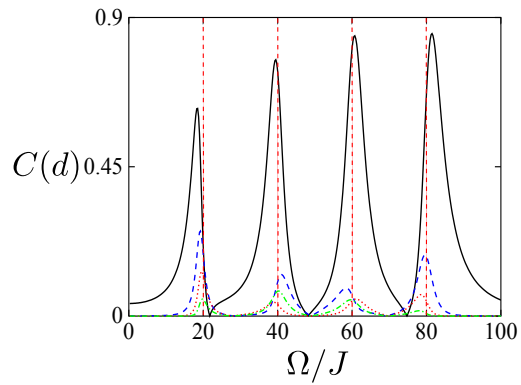}
\caption{ \label{ph_correlation} (Color online) Two-site parity correlation, ${C(d)}$ versus
coupling strength $\Omega$, for $N=M=5$, $U=20J$ and $\Delta=\epsilon=0$.  
The different distances $d$ between site $M$ and site $j$ are denoted by the 
different lines: $d=1$ (black-solid), $d=2$ (blue-dashed) and 
$d=3$ (red-dotted), and $d=4$ (green-dash-dotted) respectively.
The coupling strengths $\Omega^*_n$ are marked with
red vertical dashed lines.}
\end{figure}

In Fig.~\ref{ph_correlation}, the two-site parity
correlations $C(d)$ are plotted as a function of 
coupling strength $\Omega$. 
The sharp peaks are shown when $\Omega$ are about
$\Omega^*_n$. 
This implies that
\textit{strong particle-hole correlations will be produced
when a single atom is transported}. Also, 
the correlation function $C(d)$ decreases 
when the distance $d$ increases.

\section{Entanglement entropy}
We consider a 1D lattice to be bisected 
into two parts, i.e., the left $L$ and the right $R$ parts, respectively.
The subsystem $L$ consists of the number of $M-l$ sites from
$1,\ldots,M-l$ and the subsystem $R$
consists of the number of $l$ sites from $M-l+1,\ldots,M$. 
We study the entanglement between the atoms 
in the two parts of a lattice.

By using the Schmidt decomposition \cite{Horodecki}, the ground state $|\Psi_{\rm G}\rangle$ 
can be written as
\begin{eqnarray}
|\Psi_{\rm G}\rangle&=&\sum_i\lambda_i|\psi^i\rangle_L|\psi^i\rangle_R,
\end{eqnarray}
where $\lambda_i$ is the Schimdt coefficient.
The von-Neumann entropy \cite{Horodecki} is defined as 
\begin{eqnarray}
E(\rho)&=&-\sum_i\lambda^2_i\ln\lambda^2_i.
\end{eqnarray}

\begin{figure}[ht]
\centering
\includegraphics[height=5.0cm]{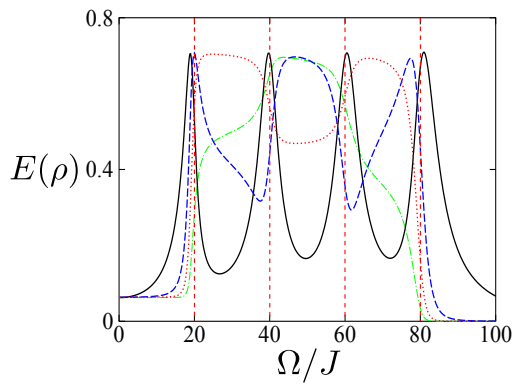}
\caption{ \label{entanglement} (Color online) Entanglement entropy, ${E(\rho)}$ versus
coupling strength $\Omega$. The same parameters in the previous figure. 
The subsystem $R$, which consists of the number $l$ of sites, is denoted by the 
different lines: $l=1$ (black-solid), $l=2$ (blue-dashed) and 
$l=3$ (red-dotted), and $l=4$ (green-dash-dotted) respectively.
The coupling strengths $\Omega^*_n$ are marked with
red vertical dashed lines.}
\end{figure}

In Fig.~\ref{entanglement}, we plot
the entanglement entropy $E(\rho)$ versus 
$\Omega$, for the different
sizes $l$. For $l=1$, the entanglement entropy 
between the particles at site $M$ and holes in the 
remaining sites is studied. 
This shows that sharp peaks occur around the 
coupling strengths $\Omega^*_n$.  This means
that a large degree of entanglement
is produced when a single atom is transported.

In fact, a large degree of 
entanglement can also be produced, for $l\geq{1}$,
as shown in Fig.~\ref{entanglement}.
For example, $E(\rho)$ rises
around $\Omega^*_n$ and becomes flat between
$\Omega^*_n$ and $\Omega^*_{n+1}$, for 
$l=3$. In the limit of large $\Omega$, 
all atoms becomes localized at site $M$.
Then, the entanglement entropy significantly 
decreases and atoms in the different sites become
unentangled in this limit.

\section{Discussion}
\begin{figure}[ht]
\centering
\includegraphics[height=5.0cm]{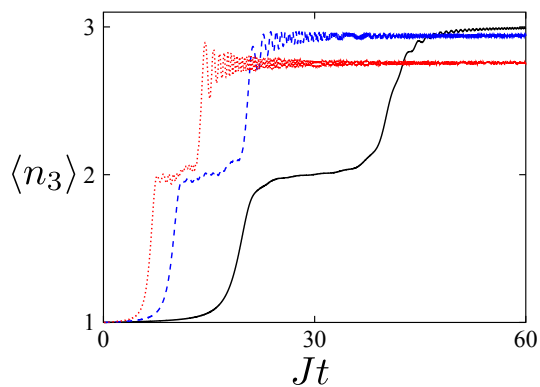}
\caption{ \label{dfig1} (Color online) Number of atoms at site 3, $\langle{n_3}\rangle$ versus
time $Jt$, for $N=M=3$, $U=20J$ and $\Delta=\epsilon=0$.  
The field is applied to the atoms at site 3 and the coupling strength 
is a linear function of time, i.e., $vt$. 
The different values of $v$, are denoted by the 
different lines: $v=J$ (black-solid), $v=2J$ (blue-dashed) and 
$v=3J$ (red-dotted) respectively.}
\end{figure}

We study the quantum transport of atoms by adiabatically
changing the coupling strength $\Omega(t)=vt$ with the time
$t$, where $v$ is a positive number. In Fig.~\ref{dfig1},
we plot the number of atoms at site $M=3$ versus time,
where the number of sites and atoms are both equal to 3.
The atoms are transported stepwise. It
is similar to the case of ground state. If the changing
rate $v$ is slow enough, then all atoms can be transferred to site 3.  
When $v$ becomes larger, the transport rate is faster. But a 
smaller number of atoms can be transported at site $M$ 
as shown in Fig.~\ref{dfig1}.
 
\begin{figure}[ht]
\centering
\includegraphics[height=5.0cm]{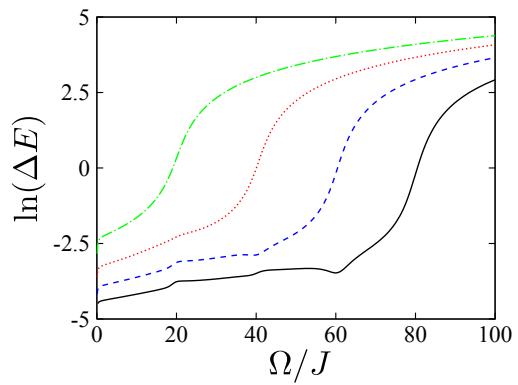}
\caption{ \label{engapfig1} (Color online) Logarithm of $\Delta{E}$ 
versus coupling strength $\Omega$, for the different numbers of atoms $N$,
and $N=M$. The different
number of $N$ are denoted by the different lines:
$N=5$ (black-solid), $N=4$ (blue-dashed), $N=3$ (red-dotted)
and $N=2$ (gree-dash-dotted), respectively. 
The parameters are: $U=20J$ and $\Delta=\epsilon=0$.  
}
\end{figure}

To implement the adiabatic transition, it is necessary
to ensure that the changing rate of 
the parameter $\Omega$ is sufficiently slow \cite{Messiah}.
It is required that the changing rate $v$
is much smaller than the energy gap $\Delta{E}$ between the first excited
state and ground state. In Fig.~\ref{engapfig1}, we plot
the logarithm of energy gap as a function of $\Omega$.
This shows that the energy gap exponentially decreases 
when the total number $N$ of atoms increases. 
Therefore, it becomes very difficult to perform the adiabatic 
transition when $N$ goes large.

\begin{figure}[ht]
\centering
\includegraphics[height=5.0cm]{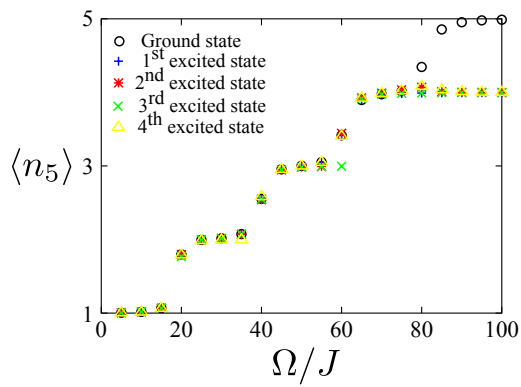}
\caption{ \label{excitfig1} (Color online) Number of atoms at site 5, $\langle{n_5}\rangle$ versus
coupling strength $\Omega$, for the first to fifth eigenstates.
The parameters are: $N=M=5$, $U=20J$ and $\Delta=\epsilon=0$.  
}
\end{figure}

To proceed, we study the transport of
atoms for the low-lying excited states.
In Fig.~\ref{excitfig1}, we plot the number of atoms at site $M$
versus $\Omega$, for the first few
eigenstates. The atoms can be
transported stepwise when $\Omega$
increases.  This method can thus be used for 
transporting a few atoms when the lattice size 
grows large. Thus, holes can be created 
in the remaining lattice sites for the low-lying
excited states. The quantum entanglement 
between the atoms and holes should be generated
in a manner similar to that in the ground state.
However, the atoms cannot all 
be transported to site $M$ for the higher excited
states.

\section{Conclusion}
In summary, we have studied the ground state
of two-component bosons in a 1D optical
lattice, where an external field is individually applied
to the atoms at one end of the lattice.  The
atoms can be transported and become localized 
at that site. In this way, the holes will be created in the remaining
sites. The particle-hole correlations can be 
produced and they can be indicated by two-site parity correlation functions.
We have also investigated the entanglement 
entropy between the atoms in the two parts of a 
lattice. The large degree of particle-hole entanglement can be 
produced in the ground state. 
We have discussed the adiabatic transport 
of the ground state and the transport properties of the
low-lying excited states.

\appendix
\section{Transport condition}
In this Appendix, we derive the transport condition of atoms
for the coupling strengths $\Omega^*_n$. 
To obtain the coupling strength $\Omega^*_n$, it is 
necessary to find out the ground-state energy of the system.
For simplicity, we consider the atoms to be trapped in the lattice
without the harmonic confinement, i.e., $\epsilon^e_i=\epsilon^g_i=0$.

In the strongly interaction regime, the number
of atoms is conserved in each site.  We write 
$S_{ix}=(e_ig^\dag_i+g_ie^\dag_i)/2$, 
$S_{iy}=(e_ig^\dag_i-g_ie^\dag_i)/2i$ and
$S_{iz}=(e^\dag_ie_i-g^\dag_ig_i)/2$.
The Hamiltonian can be written as
\begin{equation}
H=\Delta\sum_{i}\Big(S_{iz}+\frac{n_i}{2}\Big)+2\Omega_MS_{Mx}+\frac{U}{2}\sum_i{n_i(n_i-1)},
\end{equation}
where $n_i$ is the number operator at site $i$. 
This Hamiltonian can be diagonalized by applying the transformation to site $M$:
\begin{eqnarray}
S_{Mx}&=&\cos\theta{S}_{Mx}'-\sin\theta{S}_{Mz}',\\
S_{Mz}&=&\cos\theta{S}_{Mz}'+\sin\theta{S}_{Mx}',
\end{eqnarray}
and choosing  $\Delta\sin\theta + 2\Omega_m\cos\theta=0$.
The transformed Hamiltonian can be written as
\begin{eqnarray}
\label{Ground_eng1}
H'&=&\Delta\sum_{i,i\neq{M}}\Big(S_{iz}+\frac{n_i}{2}\Big)+\frac{\Delta}{2}n_M+\sqrt{\Delta^2+4\Omega^2_M}S'_{Mz}\nonumber\\
&&+\frac{U}{2}\sum_i{n_i(n_i-1)}.
\end{eqnarray}

We consider the number of
sites to be larger than or equal to the total number of atoms, i.e., 
$M\geq{N}$, then the number of atoms in site $i$ is less than one, 
where $i\neq{M}$. In this case, the ground-state energy
$E^n_G$ is 
\begin{equation}
E^n_G=\frac{\Delta{n}}{2}-\frac{n}{2}\sqrt{\Delta^2+4\Omega^2_M}+\frac{U}{2}n(n-1),
\end{equation}
where $n$ is the number of atoms in site $M$.
The coupling strength $\Omega^*_n$ can be obtained by considering $E^n_G=E^{n+1}_G$ as
\begin{eqnarray}
\Omega^*_n&=&\frac{1}{2}[(2Un+\Delta)^2-\Delta^2]^{1/2}.
\end{eqnarray}

\begin{acknowledgments}
H.T.N. thanks Jingning Zhang for useful comment.
This work was supported in part by the 
National Basic Research Program of 
China Grant 2011CBA00300, 2011CBA00302, 
the National Natural Science Foundation of 
China Grant 61033001, 61061130540.  
\end{acknowledgments}

\end{document}